\documentclass[%
onecolumn,
11pt,
nofootinbib,
amsmath,amssymb,
aps,
prd,
showpacs,
superscriptaddress
]{revtex4-1}
\usepackage{setspace}
\usepackage[caption=false]{subfig}
\usepackage{graphicx}% Include figure files
\usepackage{dcolumn}% Align table columns on decimal point
\usepackage{bm}% bold math
\usepackage{mathtools}

\DeclareMathOperator{\Tr}{Tr}
\DeclareMathOperator{\arccosh}{arccosh}
\usepackage{mathrsfs}

\usepackage{color}
\definecolor{purple}{rgb}{0.5,0,0.5}

\newcommand{\bra}[1]{\langle #1 |}
\newcommand{\ket}[1]{| #1 \rangle}

\newcommand{\ketbra}[2]{|#1\rangle \langle #2 |}

\usepackage[unicode=true,bookmarks=true,
bookmarksnumbered=false,bookmarksopen=false,
breaklinks=false,
pdfborder={0 0 1},
backref=false,colorlinks=true]{hyperref}

\hypersetup{pdftitle={De Sitter Space as a Tensor Network: Cosmic No-Hair, Complementarity, and Complexity}, pdfauthor={Ning Bao,  ChunJun Cao, Sean M. Carroll, Aidan Chatwin-Davies}, citecolor=blue,linkcolor=blue,urlcolor=blue,citecolor=blue}

\newcommand{\mrm}[1]{\mathrm{#1}}
\newcommand{\Hil}{\mathcal{H}}

\newtheorem{thm}{Theorem}[section]
\newtheorem{prop}[thm]{Proposition}

\newtheorem{defn}[thm]{Definition}

\newcommand{\pf}{\noindent {\bf Proof: }}
\newcommand{\eop}{{\hspace*{\fill}$\square$}}

\newtheorem{example}[thm]{Example}
\newenvironment{eg}{\begin{example}  \rm  }{    \end{example} }

\newcommand{\Fig}[1]{Fig.~\ref{#1}}
\newcommand{\App}[1]{App.~\ref{#1}}

\begin{document}

\preprint{CALT-TH-2017-046}

\title{De~Sitter Space as a Tensor Network: \\ Cosmic No-Hair, Complementarity, and Complexity}

\author{Ning Bao}
\email{ningbao75@gmail.com}
\affiliation{Walter Burke Institute for Theoretical Physics, California Institute of Technology, Pasadena, CA 91125, USA}
\affiliation{Institute of Quantum Information and Matter, California Institute of Technology, Pasadena, CA 91125, USA}
\author{ChunJun Cao}
\email{ccj991@gmail.com}
\affiliation{Walter Burke Institute for Theoretical Physics, California Institute of Technology, Pasadena, CA 91125, USA}
\author{Sean M.\ Carroll}
\email{seancarroll@gmail.com}
\affiliation{Walter Burke Institute for Theoretical Physics, California Institute of Technology, Pasadena, CA 91125, USA}
\author{Aidan Chatwin-Davies}
\email{achatwin@caltech.edu}
\affiliation{Walter Burke Institute for Theoretical Physics, California Institute of Technology, Pasadena, CA 91125, USA}

\begin{abstract}

We investigate the proposed connection between de~Sitter spacetime and the MERA (Multiscale Entanglement Renormalization Ansatz) tensor network, and ask what can be learned via such a construction.
We show that the quantum state obeys a cosmic no-hair theorem: the reduced density operator describing a causal patch of the MERA asymptotes to a fixed point of a quantum channel, just as spacetimes with a positive cosmological constant asymptote to de~Sitter.
The MERA is potentially compatible with a weak form of complementarity (local physics only describes single patches at a time, but the overall Hilbert space is infinite-dimensional) or, with certain specific modifications to the tensor structure, a strong form (the entire theory describes only a single patch plus its horizon, in a finite-dimensional Hilbert space). We also suggest that de~Sitter evolution has an interpretation in terms of circuit complexity, as has been conjectured for anti-de~Sitter space.

\end{abstract}

\maketitle

\baselineskip=13pt

\tableofcontents

\section{Introduction}

Even in the absence of a completely-formulated theory of quantum gravity, a great deal can be learned by combining insights from classical gravity, semiclassical entropy bounds, the principles of holography and complementarity, and the general structure of quantum mechanics. 
A natural testing ground for such ideas is de~Sitter space, a maximally symmetric spacetime featuring static causal patches with a finite entropy.
De~Sitter is also of obvious phenomenological relevance, given the positive value of the cosmological constant in the real world.
In this paper we apply ideas from quantum circuits and tensor networks to investigate quantum properties of de~Sitter on super-horizon scales.

The Multiscale Entanglement Renormalization Ansatz (MERA) is a well-studied tensor network that was originally developed to find ground states of 1+1 dimensional condensed matter theories \cite{Vidal-MERA:2008}.
In recent years, an interesting connection has been drawn between the MERA and $\mrm{AdS}_3$/$\mrm{CFT}_2$, by way of using the MERA to discretize the AdS space \cite{Swingle:2009bg,Swingle:2012wq}.
The argument was made that this could be seen as a way of emerging AdS space from the boundary CFT, thus establishing AdS/CFT as a theory in which bulk spacetime emerges from entanglement properties on the boundary.
Further work exploring this direction and generalizing it to other types of tensor networks has been done by \cite{Qi:2013caa,Pastawski:2015qua, Hayden:2016cfa}, and a $p$-adic approach to AdS/CFT using trees is explored by \cite{Heydeman:2016ldy,Gubser:2016htz}.
However, the AdS/MERA correspondence seems to have tensions with other known results in holography. For example, it is puzzling that AdS/MERA appears to suggest a ``bulk geometry'' in the form of a tensor network even for a CFT with a small central charge. Additionally, it needs to satisfy a set of stringent constraints, brought on by the fact that it is supposed to duplicate the established results of AdS/CFT \cite{Maldacena:1997re,Ryu:2006bv}.
It appears that AdS/MERA in its simplest form is not able to satisfy all of the constraints imposed by holography with AdS geometry \cite{Bao:2015uaa,Czech:2015kbp,Czech:2015xna}, although extensions may be able circumvent this difficulty \cite{Evenbly:2017}.

There is also considerable interest in studying a more general notion of geometry from entanglement beyond the context of AdS/CFT \cite{Cao:2016mst}, where geometries are related to our physical universe \cite{Strominger:2001pn,Sekino:2009kv}. 
A connection between the MERA and de~Sitter spacetime has been suggested, where we think of the tensors as describing time evolution, rather than as relating different spatial regions \cite{Beny:2013,Czech:2015kbp,SinaiKunkolienkar:2016lgg}.
In the case of 1+1 dimensions, it is also claimed \cite{Czech:2015kbp} that the MERA can be thought of as a discretization of a slice in the ``kinematic space'' \cite{Czech:2015qta,Czech:2016xec}, which corresponds to the space of geodesics in the hyperbolic plane in the particular case of $\mrm{AdS}_3/\mrm{CFT}_2$. This beautifully illustrates a correspondence between regions in the dual kinematic space, which take on information-theoretic interpretations, and the individual tensors localized in the MERA. 
More tentatively, quantum circuits have been proposed as a way of studying realistic cosmological evolution from inflation to the present epoch and beyond \cite{bao_etal2017}.

In this paper we investigate this proposed connection between the MERA and de~Sitter, under the assumption that a MERA-like circuit is able to simulate effective quantum gravitational dynamics on super-Hubble scales for some subset of quantum states in a theory of quantum gravity.
We show that the structure of the MERA is able to reproduce some desirable features of evolution in a de~Sitter background.
In particular, we identify a scale invariant past causal cone as the static patch where an analogous light-like surface functions as the cosmic horizon.
Then we show that a version of the cosmic no-hair theorem can be derived from the fixed point of the quantum channel, whereby any state will asymptote to the channel fixed point at future infinity. 
We next examine the issue of horizon complementarity in the MERA context, and argue that the global and local descriptions of de~Sitter \cite{Nomura:2011dt,Nomura:2011rb} can be equivalent up to a unitary change of basis. We observe similarities between a strong version of local de~Sitter and the implementation of a quantum error correcting code. 
Lastly, we derive a bound on the quantum complexity of the MERA circuit, and show that the complexity scales in a manner that is consistent with the ``complexity equals action'' conjecture \cite{Brown:2015bva}.

\section{The MERA and the de~Sitter causal patch}

\begin{figure}
\centering
\includegraphics[scale=0.75]{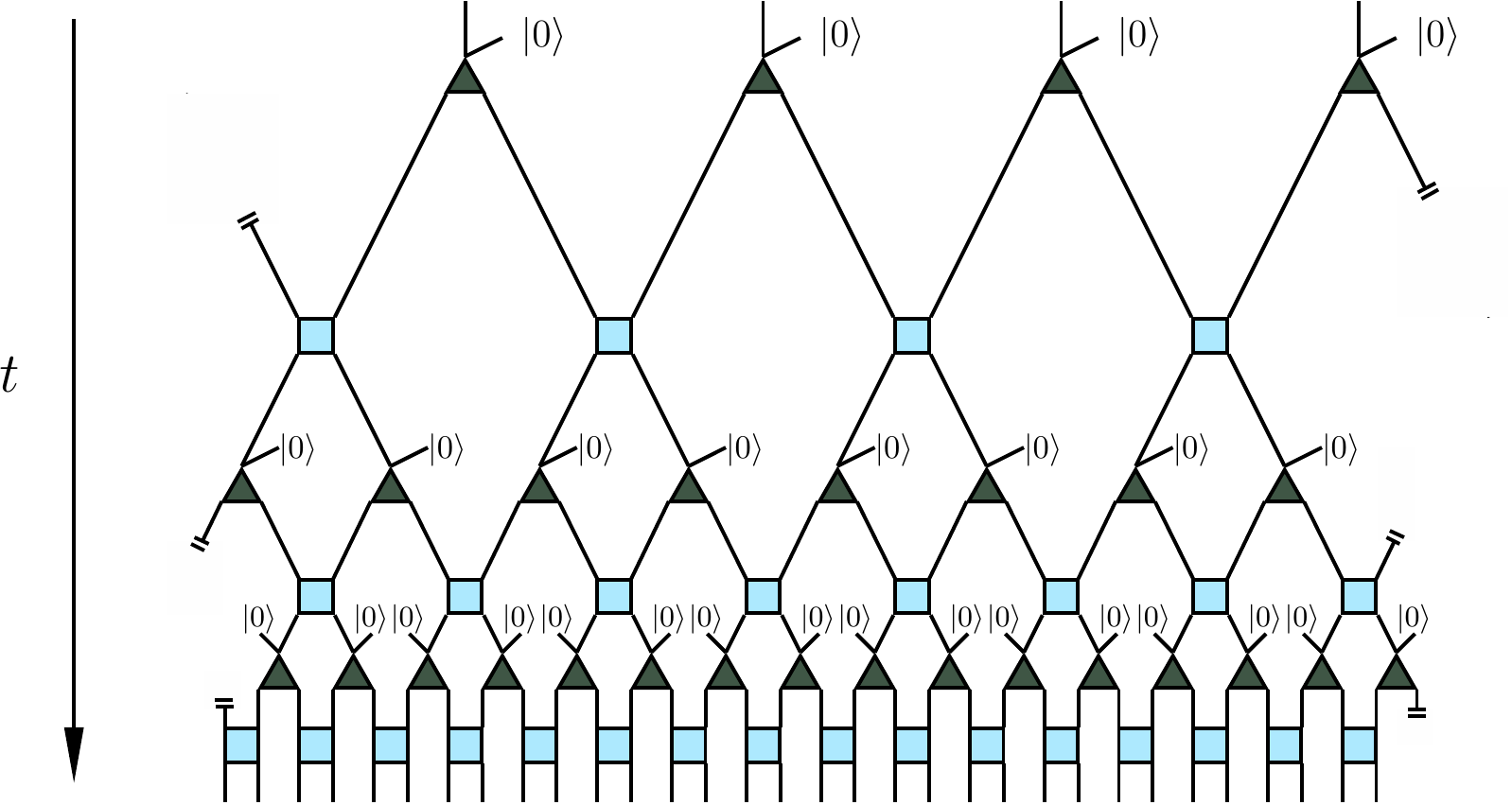}
\caption{A periodic binary MERA. The green triangles denote the isometries and the blue squares denote the disentanglers. The kets labeled $|0\rangle$ are ancilla states inserted into each isometry. The action of the circuit is to take a state at the top and evolve it downward. In anticipation of the connection to de Sitter, the fine-graining direction is labelled as the direction of increasing $t$.}
\label{fig:binaryMERA}
\end{figure}

In \Fig{fig:binaryMERA} we illustrate the MERA tensor network.
In its original conception as an ansatz for constructing ground states of 1-d spin systems, one starts with a simple quantum state at the top of the diagram, and propagates it downward through a series of gates to a final state at the bottom.
Each line represents a factor of Hilbert space, which might be quite high-dimensional.
Moving downward is the ``fine-graining'' direction, and upward is ``coarse-graining.''
The square gates are ``disentanglers'' (although they create entanglement as we flow downward), which take two factors in and output another two factors.
The triangular gates are ``isometries,'' which can be thought of as taking in a single factor and outputting two factors; alternatively, we can imagine inputting two factors, one of which is a fixed state $|0\rangle$, and outputting another two, so that the total dimensionality entering and exiting each tensor is equal. 
We will adopt the latter perspective in this paper.
It is often convenient to consider generalizations where $k>2$ factors enter and exit each tensor.

In the AdS/MERA correspondence, tensors are taken to represent factors of Hilbert space, and the two-dimensional geometry of the graph is mapped to the hyperbolic plane.
Here, where we are interested in studying a dS/MERA correspondence, flow through the circuit represents evolution through time.
Note that, while it is common in general relativity to draw spacetime diagrams with the future at the top, the convention in quantum circuits for MERA is to start with one or more ``top tensors'' and evolve downward. 
Here we will stick to the conventions of the respective communities; time flows downward in MERA circuit diagrams, and upward in spacetime diagrams.\footnote{We will occasionally draw circuit diagrams in which time flows from left to right, just to keep things lively.}

In this work, we will mostly be concerned with MERAs that are scale and translationally invariant (the same disentanglers and isometries appear everywhere in the network).
We use the term ``site'' in the MERA to refer to a Hilbert space factor that lives on a leg that exits a disentangler (or equivalently, that enters an isometry).
When the MERA is used as a variational ansatz for a physical system like a spin chain, the collection of sites at any given layer corresponds to the state of the physical lattice at that renormalization scale.
For more extensive reviews of tensor networks and the MERA see \cite{Evenbly:2007hxg,Evenbly:2011,Bao:2015uaa}.

Viewed as a circuit in which the fine-graining direction corresponds to the future or past direction (away from the de~Sitter throat), the MERA reproduces the causal structure of de~Sitter spacetime \cite{Beny:2013,Czech:2015kbp,SinaiKunkolienkar:2016lgg}.
Recently, as a part of their studies of kinematic space, Czech \emph{et al.} further pointed out that there is a natural way of associating the MERA with half of the 1+1-dimensional de~Sitter manifold \cite{Czech:2015kbp}.
Here we briefly explain how this works.

\begin{figure}[b]
\centering
\includegraphics[scale=0.75]{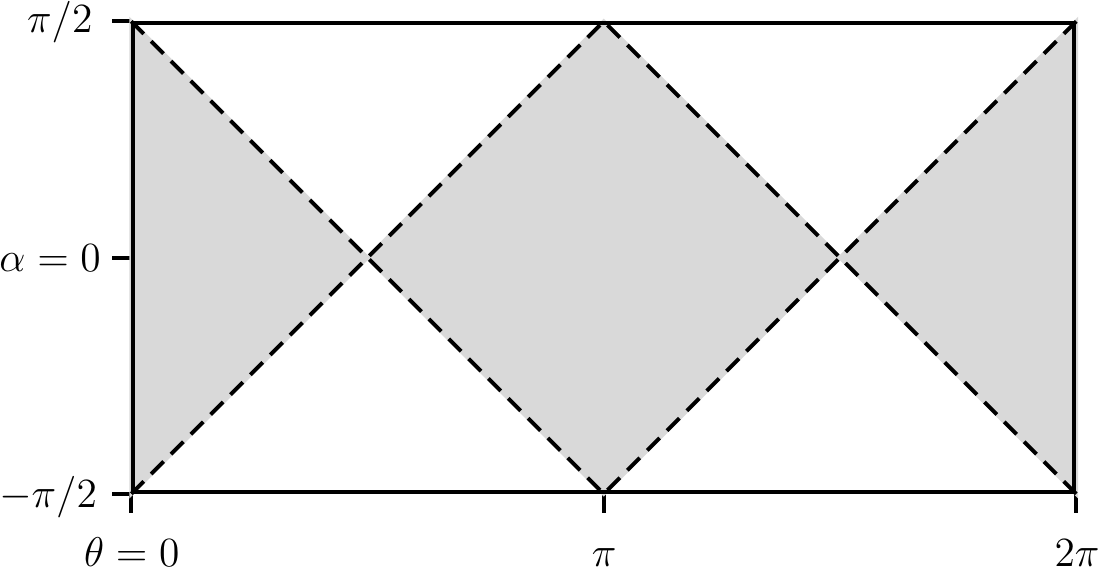}
\caption{The Penrose diagram of global (1+1)-dimensional de~Sitter spacetime. As this is a spacetime diagram, time now runs from bottom to top. The boundaries of two complete disjoint causal patches, one centered at $\theta = 0$ and the other centered at $\theta = \pi$, are drawn with a dashed line, and the interiors of the patches are shaded. Light rays travel along $45^\circ$ lines in this diagram.}
\label{fig:dSpenrose}
\end{figure}

Let $\mathcal{M}$ be 1+1-dimensional de~Sitter spacetime with the usual global coordinatization:
\begin{equation}
ds^2 = \ell_\mrm{dS}^2(-dt^2+\cosh^2 t \, d\theta^2).
\end{equation}
The timelike coordinate $t$ takes all real values, and $\theta$ is an angular coordinate that is $2\pi$-periodic.
In these coordinates, $\mathcal{M}$ looks like a hyperboloid whose constant-$t$ sections are circles that attain a minimum radius at $t=0$ and that grow in either direction away from $t=0$.
The proper radius at $t=0$ is equal to $\ell_\mrm{dS}$, which is called the de~Sitter radius.
A convenient coordinate transformation is to set $\cosh t = \sec \alpha$, under which the metric becomes conformally flat:
\begin{equation}
ds^2 = \frac{\ell_\mrm{dS}^2}{\cos^2 \alpha} \left( -d\alpha^2 + d\theta^2 \right).
\end{equation}
Because of this, the full de~Sitter manifold is often represented by a rectangle in the $\theta$-$\alpha$ plane with $-\pi/2 < \alpha < \pi/2$ and $0 < \theta < 2\pi$, as in the Penrose diagram of de~Sitter, \Fig{fig:dSpenrose}.

\begin{figure}[t]
\centering
\includegraphics[width=\textwidth]{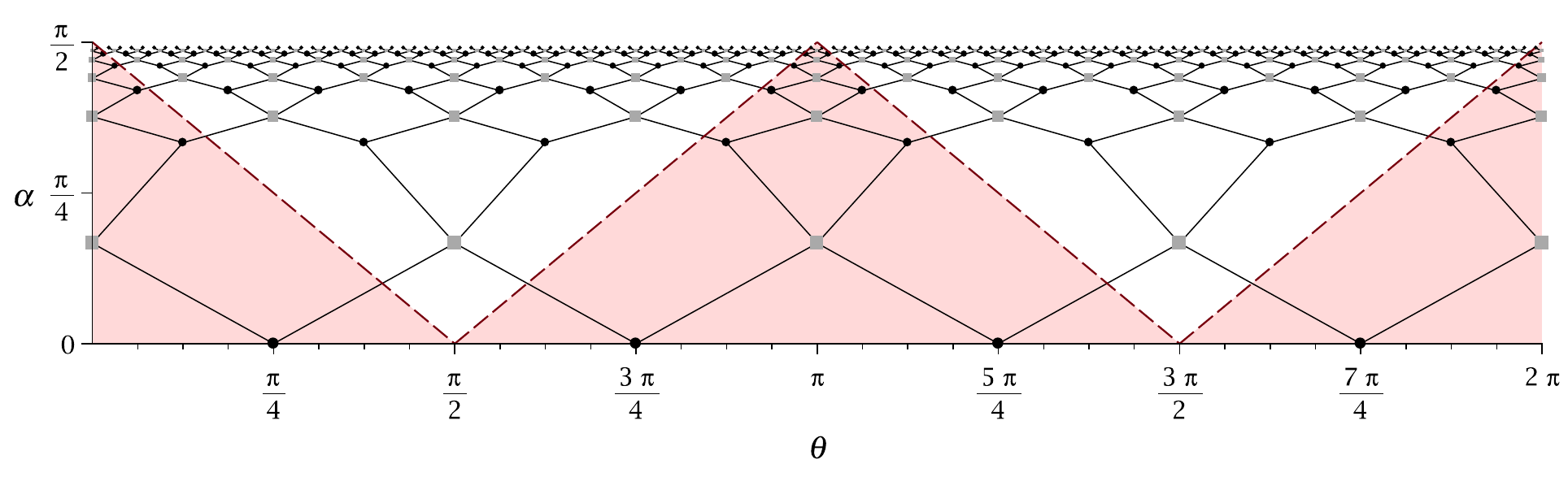}
\caption{A geometric de~Sitter-MERA correspondence, mapping the MERA circuit to the top half of the de~Sitter geometry. Note that the fine-graining direction of the MERA in this diagram points upward to match the future direction in the Penrose diagram. The domain of dependence of any pair of adjacent sites in the initial layer of the MERA is entirely contained within a single static patch in de~Sitter. Two of the four possible static patch interiors are shaded in red. (The other two static patches are centered at $\theta = \pi/2$ and $\theta = 3\pi/2$.)}
\label{fig:dS-MERA}
\end{figure}

Consider now the top half of the de~Sitter manifold with $t \geq 0$ (or $0 \leq \alpha < \pi/2$).
Starting at $t_0 \equiv 0$, the length of the constant-$t_n$ slice doubles
at every subsequent time $t_n = \arccosh 2^n$ with $n = 1, 2, \dots$
This suggests identifying the top of a translationally invariant binary MERA with the $t_0 = 0$ slice, and subsequent layers of the MERA with the subsequent $t_n$ slices, so that the MERA describes the top half of the de~Sitter hyperboloid. 
This identification is illustrated in \Fig{fig:dS-MERA}, in which the sites of the $n^\mrm{th}$ layer of the MERA have been chosen to lie at the angles
\begin{equation}
\theta_j^{(n)} = \frac{\pi}{2^{n+1}} \left(j + \frac{1}{2} \right) \qquad j = 0, \, \dots \, , 2^{n+2}-1 \, .
\end{equation}

The fact that the top of the MERA was chosen to have four sites was no coincidence.
With this choice, the \emph{future domain of dependence} of any two adjacent sites at the top of the MERA precisely coincides with (the top half of) a single static patch of de~Sitter.
Or, to use terminology that is more familiar in the MERA literature, each static patch of de~Sitter that is centered at $\theta = 0$, $\pi/2$, $\pi$, or $3\pi/2$ coincides with a causal cone \cite{Evenbly2014} in the MERA such that every layer of the causal cone contains precisely two sites of the MERA (i.e., the causal cone is stationary).

Let us elaborate a bit on the terminology above.
First, recall how a domain of dependence is defined on a smooth manifold:
\begin{defn}
Let $S \subset \mathcal{M}$ be a subset of a smooth Lorentzian manifold $\mathcal{M}$.
The future (resp. past) domain of dependence of $S$ is the set of all points $p \in \mathcal{M}$ such that every past (resp. future) inextensible causal curve through $p$ intersects $S$.
\end{defn}
This suggests the following analogous definition for a domain of dependence in a MERA:
\begin{defn}
Let $S$ be a collection of sites in a MERA.
The future (resp. past) domain of dependence of $S$ is the set of all MERA sites $p$ such that starting at $p$ and moving only in the past, or coarse-graining direction (resp future, or fine-graining direction), one inevitably arrives at a site in $S$. 
\end{defn}
In de~Sitter space, the proper radius of the cosmological horizon is constant.
Given an inextendible timelike geodesic, a static patch is defined as the set of all points connected to that geodesic by both past- and future-oriented causal curves, and its size is given by the horizon radius.
In particular, in 1+1 dimensions the horizon radius is $\pi \ell_\mrm{dS}/2$.
Within a constant-$t$ slice, a horizon volume is an interval of proper length $\pi \ell_\mrm{dS}$, and static patches are diamonds in the Penrose diagram (cf. \Fig{fig:dSpenrose}).

In line with \cite{Czech:2015kbp}, we here adopt a correspondence between the MERA and half of the full 1+1-dimensional de~Sitter manifold in which stationary causal cones in the MERA are in correspondence with static patches of de~Sitter.
In the spirit of tensor-network/spacetime correspondences, one should think of the MERA and the state that it describes as some state of quantum gravity describing quantum fields evolving in a semiclassical de~Sitter background. 
In other words, despite lacking an explicit theory of quantum gravity, we suggest that some aspects of the effective dynamics for a quantum gravity state that describes classical de~Sitter spacetime can be described and organized at a fundamental level by a suitably-chosen MERA.

In this picture, each site of the MERA carries a Hilbert space $\Hil_*$, and the Hilbert space that corresponds to a given horizon volume, call it $\Hil_\mrm{static}$, is the tensor product of the Hilbert spaces of the sites that lie within the horizon.
We do not count the Hilbert spaces that correspond to unentangled ancillae as part of the static patch Hilbert space, since we only attach a spacetime interpretation to entangled degrees of freedom in the MERA proper.
To be consistent with the Gibbons-Hawking entropy of de~Sitter spacetime \cite{Gibbons1977}, it should be that $\ln \dim \mathcal{H}_{\rm static} \sim S_{\rm dS}$, where $S_{\rm dS}$ is the de~Sitter entropy.
Hence, for our Universe, where $S_{\rm dS} \sim 10^{122}$, the corresponding bond dimension (i.e., the dimensionality of $\Hil_*$) is of order $\dim \Hil_* \sim \exp(10^{122})$ per site.

This is a very coarse-grained description of de~Sitter spacetime. For a binary MERA, there are only two sites per horizon volume, and layers of the MERA within a static patch are separated by cosmological timescales.
Furthermore, a binary MERA only accommodates 4 distinct static patches (\Fig{fig:dSpenrose}).
We imagine, however, that it should be possible to refine this horizon-scale description via, e.g., local gadget expansions, in which the large Hilbert space $\Hil_*$ could be factorized according to sub-horizon locality.
This perhaps can be achieved by some version of cMERA \cite{Haegeman:2011uy,Nozaki:2012zj,Miyaji:2016mxg}.

\begin{figure}
\centering
\includegraphics[width=\textwidth]{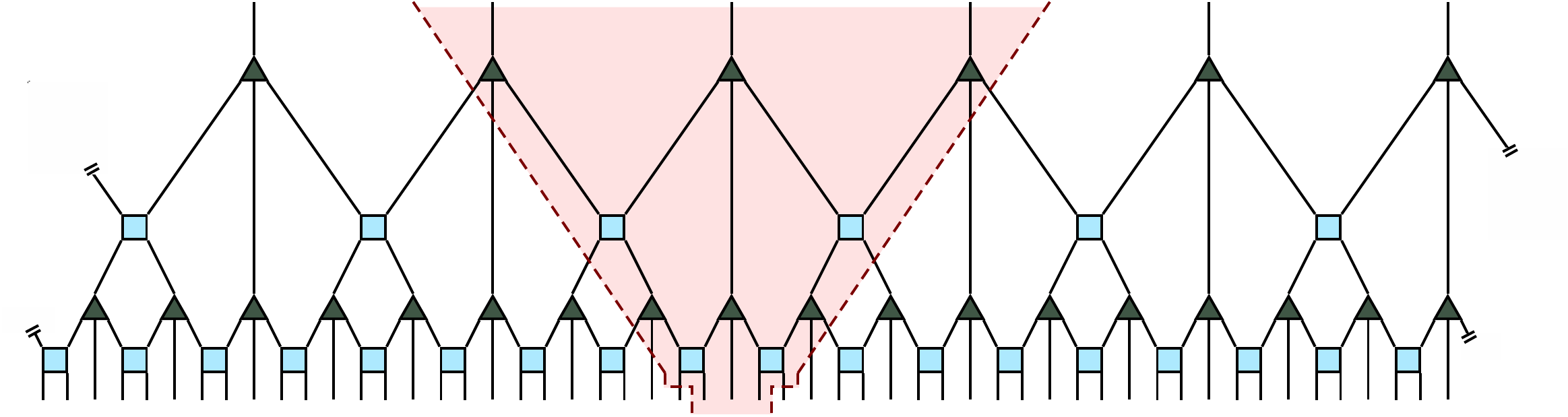}
\caption{A ternary MERA. Ancillae are suppressed in this diagram. A stationary causal cone with three sites per layer is indicated by the shaded region.}
\label{fig:ternaryMERA}
\end{figure}

One might wonder whether it is possible to pack more MERA sites into a single slice of the static patch by starting with more sites at the top of the MERA, or by considering a MERA with a larger branching factor.
The number of sites at the top of the MERA is fixed by the number of sites per layer in the stationary causal cone, however.
If the stationary causal cone has $m$ sites per layer, then the $t=0$ slice contains $2m$ sites.
The reason is simply because the $t=0$ slice of de~Sitter contains exactly two disjoint horizon volumes.
The quantity $m$ in turn is fixed by the branching factor and the structure of the MERA.
For a binary MERA, a stationary causal cone always has $m=2$ sites per layer.
A ternary MERA has $m=3$ sites per layer in a stationary causal cone (\Fig{fig:ternaryMERA}).
However, in general for a $k$-nary MERA, in which the number of sites increases $k$-fold in each layer of the MERA, there can only ever be $m=2~\mrm{or}~3$ sites per layer in a stationary causal cone.
Further details of stationary causal cones and a proof of this last fact are given in \App{app:stationary}.

Unfortunately, the global de~Sitter-MERA correspondence as formulated on a (hyper)cubic lattice does not easily generalize to higher dimensions due to discretization artifacts.
The possibility of a de~Sitter-MERA correspondence in higher dimensions is discussed in \App{app:MOREDIMS}.

\section{Cosmic No-Hair as a channel property}

\begin{figure}[ht]
\centering
\subfloat[]{
\includegraphics[scale=0.45,width=0.4\textwidth]{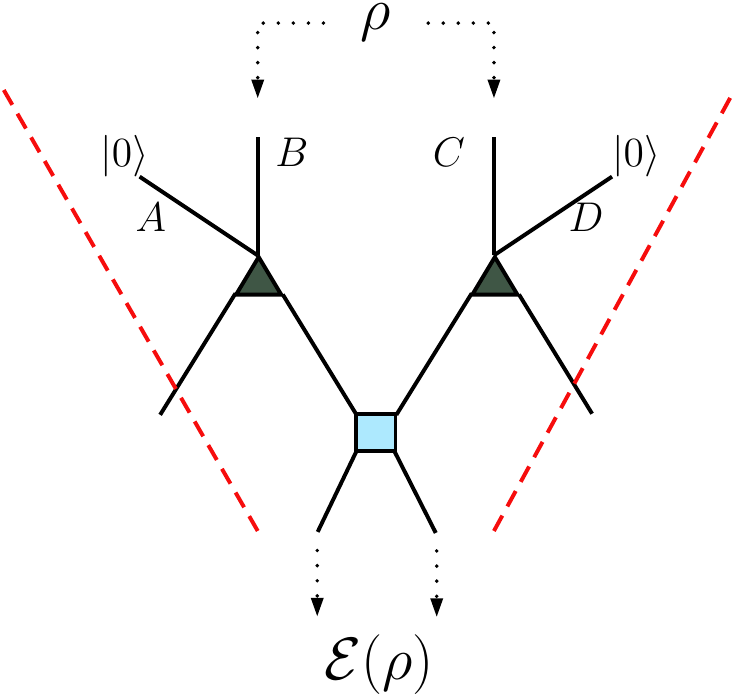}
}
\subfloat[]{
\includegraphics[scale=0.45,width=0.4\textwidth]{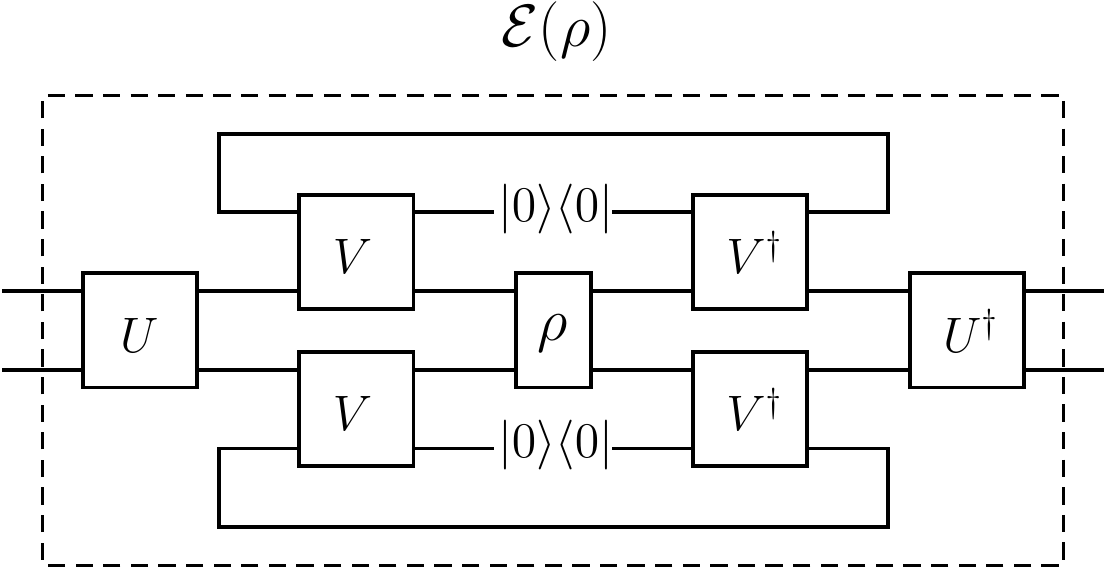}
}
\caption{(a) A single step of the MERA within the causal patch, viewed as a channel $\mathcal{E}$, and (b) the equivalent circuit diagram. 
Time runs in the downward direction in (a).}
\label{fig:stepchannel}
\end{figure}

Via the correspondence described above, each constant-$t$ slice of a de~Sitter static patch is assigned a Hilbert space
\begin{equation}
\Hil_\mrm{static} = \Hil_* \otimes \Hil_*,
\end{equation}
where $\Hil_*$ is the Hilbert space of a single MERA site.
If we restrict our attention to a single static patch, then the MERA also defines a superoperator, $\mathcal{E}$, which maps a state in $\Hil_\mrm{static}$ forward by one Hubble time to a state on the next slice.
With the disentanglers and isometries held fixed and uniform across the MERA, the action of $\mathcal{E}$ may be written explicitly as
\begin{equation}
\mathcal{E}(\rho) = U_{BC} \, \Tr_{AD} \left[ V_{AB} \otimes V_{CD} (\ket{0} \bra{0}_A \otimes \rho_{BC} \otimes \ket{0}\bra{0}_D) V_{AB}^\dagger \otimes V_{CD}^\dagger  \right] U_{BC}^\dagger.
\end{equation}
The labels $A$, $B$, $C$, and $D$ indicate on which Hilbert space factors operators act, but we may subsequently omit them when it does not cause confusion.
The ancillae are labelled by $A$ and $D$, and $\Hil_\mrm{static}$ is labelled by $B$ and $C$, cf. \Fig{fig:stepchannel}.

In the MERA literature, $\mathcal{E}$ is known as the \emph{descending superoperator} \cite{Evenbly:2007hxg,Pfeifer:2008}.
It is a quantum channel by construction, i.e., it is completely positive and trace-preserving on the set of states (density operators), which for future reference we will denote by $\mathcal{S}(\Hil_\mrm{static})$.
In precise language, given a Hilbert space $\Hil$, if $\mathscr{H}(\Hil)$ denotes the space of Hermitian operators on $\Hil$, then the set of states is
\begin{equation}
\mathcal{S}(\Hil) \equiv \left\{ \rho \in \mathscr{H}(\Hil) \; | \; \Tr~\rho = 1, \; \bra{\psi} \rho \ket{\psi} \geq 0 ~~ \forall ~ \ket{\psi} \in \Hil \right\} .
\end{equation}

Consider now starting at some given layer with a state $\rho_0 \in \mathcal{S}(\Hil_\mrm{static})$ and repeatedly applying the map $\mathcal{E}$.
Intuitively, every application of $\mathcal{E}$ dilutes the original state $\rho_0$ by entangling it with the same ancillary state $\ket{00}\bra{00}_{AD}$ before taking a partial trace, at which point information about $\rho_0$ flows out of the static patch. 
It is therefore natural (and correct) to expect that the state on the static patch should settle down to a future asymptotic steady state, regardless of the initial state $\rho_0$.

We will make this expectation rigorous below, but first we note that this observation suggests a sort of \emph{cosmic no-hair theorem} for the de~Sitter-MERA correspondence.
In classical general relativity, a cosmic no-hair theorem is roughly the statement that a positive cosmological constant causes a  spacetime to asymptotically tend to a de~Sitter state in the future.
The following theorem of Wald pertaining to Bianchi spacetimes, which are homogeneous but anisotropic cosmological models, is perhaps the most precise statement of a cosmic no-hair theorem \cite{Wald:1983ky}:
\begin{thm}[Wald] All Bianchi spacetimes (with the exception of certain strongly-curved Bianchi IX spacetimes) that are initially expanding, that have a positive cosmological constant, and whose matter content obeys the strong and dominant energy conditions asymptote to de~Sitter in the future.
\end{thm}
Various generalizations and variations of this theorem exist in the literature \cite{Starobinskii1983,Barrow1987,Barrow1989,Kitada1992,Kitada1993,Bruni1995, Bruni2002, Boucher2011, Maleknejad2012, Carroll:2017kjo}.
In particular, quantum cosmic no-hair theorems show that the quantum states of fields tend to their respective vacuum states on an asymptotically de~Sitter background \cite{Hollands2010,Marolf2010,Marolf2011}.
The MERA results here are reminiscent of these quantum cosmic no-hair theorems.

Let us now add some rigor to the above observations.
When $\Hil$ is finite-dimensional, quantum channels are necessarily contractions on $\mathcal{S}(\Hil)$ \cite{Raginsky:2002}.
Recall that a linear map $T : X \rightarrow X$ on a Banach space $X$ is a \emph{contraction} if there exists $0 < \kappa \leq 1$ such that $d(T(x_1),T(x_2)) \leq \kappa \, d(x_1,x_2)$ for all $x_1, x_2 \in X$, where $d$ is the metric on $X$.
For $\mathcal{S}(\Hil)$, the metric is most commonly defined using the 1-norm,
\begin{equation}
d(\rho,\sigma) \equiv || \rho - \sigma ||_1 ,
\end{equation}
where $|| A ||_1 = \Tr \sqrt{A^\dagger A}$ for any linear operator $A$.\footnote{All norms are equivalent in finite dimensions, i.e., for any two norms $\Vert \cdot \Vert_a$ and $\Vert \cdot \Vert_b$, there exist constants $m>0$ and $M>0$ such that $m \Vert v \Vert_a \leq \Vert v \Vert_b \leq M \Vert v \Vert_a$ for all $v$ in the normed space.}
A contraction is \emph{strict} when $0 < \kappa < 1$, in which case the contraction mapping principle guarantees that there is a unique fixed point $x_\star \in X$ such that $T(x_\star) = x_\star$.
Furthermore, the sequence $\{T^n(x_0)\}_{n=1}^{\infty}$ converges to the fixed point $x_\star$ for any choice of the starting point $x_0$.

Quantum channels need not be strict contractions in general; however, it is certainly easy to write down channels that are strict contractions \cite{Raginsky:2002}.
Returning to the de~Sitter-MERA correspondence, we may simply suppose that the disentanglers $U$ and isometries $V$ are chosen such that the superoperator $\mathcal{E}$ is a strict contraction.
Moreover, numerical assays seem to indicate that this is generally the case for random $U$ and $V$ \cite{Pfeifer:2008,Evenbly:2007hxg}.
Our intuition that the state in a causal patch should tend to some asymptotic fixed state in the future is therefore warranted.

Regardless of the channel's contractive properties, it is easy to see that $\mathcal{E}$ has at least one fixed point by examining its adjoint.
To define the adjoint, take the domain of $\mathcal{E}$ to be the space of Hermitian operators, $\mathscr{H}(\Hil)$, which is closed under addition and multiplication by real numbers.
The space $\mathscr{H}(\Hil)$ with the Frobenius inner product
\begin{equation}
\langle T, S \rangle \equiv \Tr \left(S^\dagger T \right)
\end{equation}
is then a Hilbert space over the real numbers.
As usual, the adjoint operator is defined by the relation $\langle \mathcal{E}(T), S \rangle = \langle T, \mathcal{E}^\dagger(S) \rangle$.
Using this definition, it is straightforward to show that the action of $\mathcal{E}^\dagger$ is
\begin{equation}
\mathcal{E}^\dagger (S) = {}_{AD}\bra{00} V^\dagger_{AB} V^\dagger_{CD} \left[ I_{AD} \otimes (U^\dagger S U)_{BC} \right] V_{AB} V_{CD} \ket{00}_{AD} \, .
\end{equation}
In the MERA literature, $\mathcal{E}^\dagger$ is known as the \emph{ascending superoperator}.
In this form, it is clear that the identity operator is an eigenvector of $\mathcal{E}^\dagger$ with eigenvalue $\lambda = 1$.
Therefore, $\bar{\lambda} = \lambda = 1$ is also in the spectrum of $\mathcal{E}$, or in other words, $\mathcal{E}$ necessarily has a fixed point.

That $\lambda = 1$ is an eigenvalue of $\mathcal{E}$ is well-known \cite{Evenbly:2007hxg,Pfeifer:2008}; however, we exhibited $\mathcal{E}^\dagger$ because it clearly shows that, in general, $\mathcal{E}$ is not self-adjoint.
In particular, this means that the eigenvector of $\mathcal{E}$ to the eigenvalue 1, call it $\rho_\star$, is not trivially the identity operator.
An interesting question is how much freedom is possible in choosing $\rho_\star$ by specifying the disentanglers and isometries $U$ and $V$.
Clearly there are families fixed points.
For example, if $\rho_\star$ is such that $\mathcal{E}(\rho_\star) = \rho_\star$ for a given choice of $U$ and $V$, then $\tilde{\rho}_\star \equiv (W^\dagger \otimes W^\dagger) \rho_\star (W\otimes W)$ is the fixed point of the channel $\tilde{\mathcal{E}}$ with $\tilde U = W^\dagger U$ and $\tilde V = (I \otimes W) V$ for any unitary operator $W$ on $\Hil_*$.
From exactly what subset of $\mathcal{S}(\Hil)$ the fixed point $\rho_\star$ may be chosen is an open problem.

\section{Global de~Sitter and Complementarity}

In classical general relativity, there are no barriers to describing de~Sitter spacetime in a global way.
However, in light of complementarity \cite{Susskind:1993if}, an interesting question is whether quantum gravity also accommodates a global description of de~Sitter, or whether a fully quantum theory only exists on a single causal patch.
We will suggest that a local picture (describing only a single patch) is possible via the MERA if the Hamiltonian is essentially time-dependent; as a result, this perspective also avoids Poincar\'e recurrences. 

Complementarity, as it was originally envisioned for black holes, asserts that the ability of an observer to describe the region around them in terms of local quantum field theory on a smooth spacetime background does not extend into the unobservable region behind a horizon.
For example, when describing physics outside of the black hole in a black hole spacetime, one should think of all of the black hole's degrees of freedom as residing just above its apparent horizon on a stretched horizon \cite{Thorne:1986iy}.
Nevertheless (and neglecting possible issues regarding firewalls \cite{Almheiri:2012rt}), there should also exist a complementary description of the black hole that is appropriate to, e.g., an observer who crosses the horizon, where the black hole interior is very much a real place.
Any possible discrepancies in these two descriptions are then purportedly resolved by the fact that an observer who crosses the horizon becomes causally disconnected from the black hole exterior, and so information about these discrepancies cannot be communicated to the exterior.
Applied to de~Sitter cosmology, horizon complementarity suggests that a single observer can only describe physics using local quantum field theory in a region that stretches out to the horizon, but no farther.
To this observer, the only sign of the rest of the universe is encoded on a stretched horizon.
If one considers two observers that have overlapping horizon volumes, then there is presumably some partial mapping between their respective local descriptions of physics.

The question then arises as to whether an infinitely big spacetime outside the de~Sitter horizon actually exists in this picture.
A weak version of complementarity might posit that it does, but that its existence cannot be described by any one observer; the underlying quantum theory would nevertheless still describe states in an infinite-dimensional Hilbert space.
A stronger version would postulate that the entire quantum theory has a finite-dimensional Hilbert space (with dimension of order $e^{S_{\rm dS}}$), and all that exists can be described by a single Hubble patch and its horizon \cite{Fischler2000,Banks2000,bousso2000,Banks:2000fe,Witten:2001kn,Dyson2002,Parikh:2004wh,Nomura:2011dt,Nomura:2011rb}.
The descriptions of physics in different horizon volumes contained in different causal patches are then related by a global unitary transformation.
The distinction might seem academic, but is actually crucial: unitary evolution with a time-independent Hamiltonian in a finite-dimensional Hilbert space leads to Poincar\'e recurrences and Boltzmann brains \cite{Dyson2002,Albrecht:2004ke,Carroll:2017gkl}, which can be avoided if Hilbert space is infinite-dimensional \cite{Boddy:2014eba}.

Let us refer to the weak complementarity perspective as the ``global'' view (different regions of the classical de~Sitter spacetime have an independent existence, and Hilbert space is infinite-dimensional), and the strong complementarity perspective as the ``local'' view (there is only one patch worth of information, and Hilbert space is finite-dimensional).
The MERA tensor network, we will argue, can accommodate the local description, and with a bit of modification, the global description as well.
We find that there is a natural sense in which the information associated with any single static patch can be localized on the static patch and its horizon.
We then propose a modified network that we call SCMERA (``Strong Complementarity MERA'') that could, in principle, capture the local strong complementarity view.
In order to have consistent time-evolution in the SCMERA, we will see that it is effectively generated by a time-dependent Hamiltonian, i.e., the unitary operator that maps a layer in the SCMERA to the next layer changes as a function of depth in the network.
While such evolution is in tension with our expectations in cosmology, where the Hamiltonian evolution should be time-independent, it does avoid certain undesirable phenomena like Poincar\'e recurrences.
Given how little we know about quantum cosmology, it seems worth keeping different perspectives in mind.

\subsection{Slicing, weak complementarity and pseudo-holography}

\begin{figure}[h]
 \centering
 \includegraphics[width=0.75\textwidth]{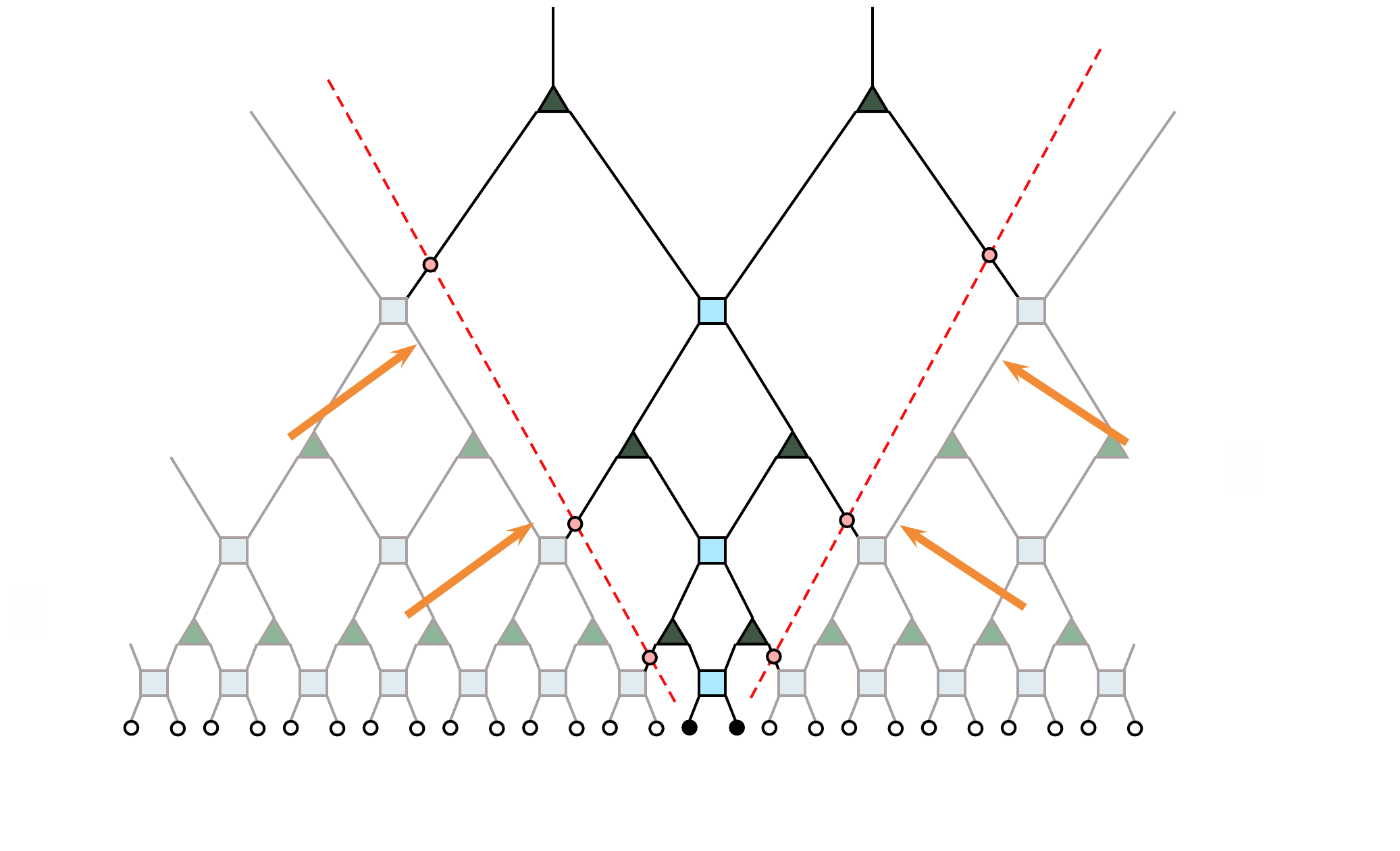}
 \caption{Sites outside the horizon at any given layer (indicated by white dots) are unitarily related, via the MERA, to a state on the horizon (indicated by red dots) and a collection of ancillae (not shown), $\ket{\tilde \Psi}_\mrm{dS} \otimes \ket{0}^{\otimes n}$. A state $|\Psi\rangle_{\rm dS}$ corresponding to the de~Sitter spatial slice is prepared at the bottom layer. The sites inside the static patch are indicated by the filled black dots.}
 \label{fig:comp_bousso}
 \end{figure}

A notable feature of the MERA is that it naturally provides a way to both define different Cauchy slices and relate the states defined on them.
Up until now, we have thought of states in global de Sitter as being defined on constant time slices, or in other words, on a single layer at constant depth in the MERA.
However, given such a state that we label by $\ket{\Psi}_\mrm{dS}$, by picking some collection of sites on which it is defined, one can define a new state $\ket{\tilde \Psi}_\mrm{dS}$ (which is in a tensor product with some collection of $n$ ancillae) and a new Cauchy slice by pushing the state on the chosen sites back up (i.e., backwards in time) through the MERA.
In other words, $\ket{\Psi}_\mrm{dS}$ and $\ket{\tilde \Psi}_\mrm{dS} \otimes \ket{0}^{\otimes n}$ are related by partial unitary evolution, and the horizontal cut through the MERA on which $\ket{\tilde \Psi}_\mrm{dS} \otimes \ket{0}^{\otimes n}$ is defined constitutes a new Cauchy slice.
In particular, given a static patch, the state $\ket{\Psi}_\mrm{dS}$ can be pushed back up through the MERA in this way so that the resulting state is supported entirely on the sites that comprise $\Hil_\mrm{static}$ and sites that are on the lightlike horizon, as illustrated in \Fig{fig:comp_bousso}.
Note that this wouldn't be possible for a generic state living on a constant $t=T$ slice in the Hilbert space of the complete theory, but can be done for the specific states that arise via the MERA from the initial state at $t=0$ (the top tensors).

The observation above suggests a toy model for weak complementarity as well as a sort of ``pseudo-holography.''
The network clearly admits a global de~Sitter description on constant time slices, but a more observer-centric view of the local patch consists of the state defined on $\Hil_\mrm{static}$ and a collection of horizon sites, as discussed above and shown in Figure \ref{fig:comp_bousso}.
For a stationary observer $\mathcal{O}_A$ who travels along a timelike geodesic at the center of the static patch, all information relevant to $\mathcal{O}_A$'s local description of physics is given by the degrees of freedom in the static patch interior.
The information about the exterior is encoded in the degrees of freedom that reside on the horizon.
However, for another observer $\mathcal{O}_B$ who travels away from $\mathcal{O}_A$ and leaves the patch, their surrounding spacetime geometry and description of the quantum state can be ``manufactured'' by propagating $\mathcal{O}_A$'s horizon degrees of freedom down through the MERA.
In this way, the region that is accessible to $\mathcal{O}_B$ is realized by decompressing \cite{Czech:2015kbp} the information that is contained on $\mathcal{O}_A$'s horizon.
The information that was previously understood to have localized on the horizon for $\mathcal{O}_A$ is, up to inclusion of ancillae, unitarily transformed to a state defined on spacetime that is to the exterior of $\mathcal{O}_A$'s static patch.
This map between the local descriptions of different observers is a realization of weak complementarity, with information about spacetime to the exterior of an observer's cosmic horizon being encoded on the horizon in a way that seems holographic.

This picture of weak complementarity is not really holographic, however, because the number of apparent degrees of freedom associated with the horizon increases toward the future in the MERA, i.e., the number of horizon sites grows with every subsequent layer.
In a true holographic model, the size of the boundary Hilbert space should remain constant.
We investigate this possibility, or in other words, the possibility of strong complementarity, in the next section.

\subsection{Strong Complementarity, recoverability and quantum error correction}

In the local, strong complementarity picture, the degrees of freedom represented by the static patch of a single observer, plus those on the corresponding horizon, together describe a closed system constituting the entirety of Hilbert space, which is correspondingly finite-dimensional. 
Ordinarily, assuming a time-independent Hamiltonian, such a setup would lead to recurrences and Boltzmann brains.
What we will find, however, is that it is more natural from the MERA perspective to imagine evolution inside the patch that is equivalent to a time-dependent Hamiltonian. (Cosmological evolution with a time-dependent Hamiltonian also plays a role in Banks and Fischler's approach to holographic spacetime \cite{Banks:2003ta,Banks:2011av,Banks:2015iya}.)

A local picture is possible in the MERA because of its particular circuit construction that begins with a finite number of inputs (4 for a binary MERA), where only two non-overlapping static patches at $t=0$ are present.
Consequently, the total number of quantum degrees of freedom for the input is limited to that of two non-overlapping patches and is, of course, finite. 
Let $\chi_* \equiv \dim \Hil_*$ denote the dimension of the Hilbert space of a single MERA site (the bond dimension).
Then, even though the number of sites in the MERA grows as a function of depth, the global state at any given subsequent layer of the MERA only resides in a subspace of dimension $\chi_*^4$.
Because $\dim \Hil_\mrm{static} = \chi_*^2$ remains the same at every step in the MERA within the static patch, there always exists a purification of the state $\rho_\mrm{static} \in \mathcal{S}(\Hil_\mrm{static})$ in a Hilbert space with dimension $\chi_*^2$.
Therefore, simply by counting Hilbert space dimensions, we could imagine that such a purifying Hilbert space, call it $\Hil_\mathrm{horizon}$, resides on the horizon of the static patch.
The horizon state would have to be unitarily related to the global state of the MERA outside the static patch (which is a preferred purification of $\rho_\mrm{static}$).

\begin{figure}[h]
\centering
\includegraphics[width=0.75\textwidth]{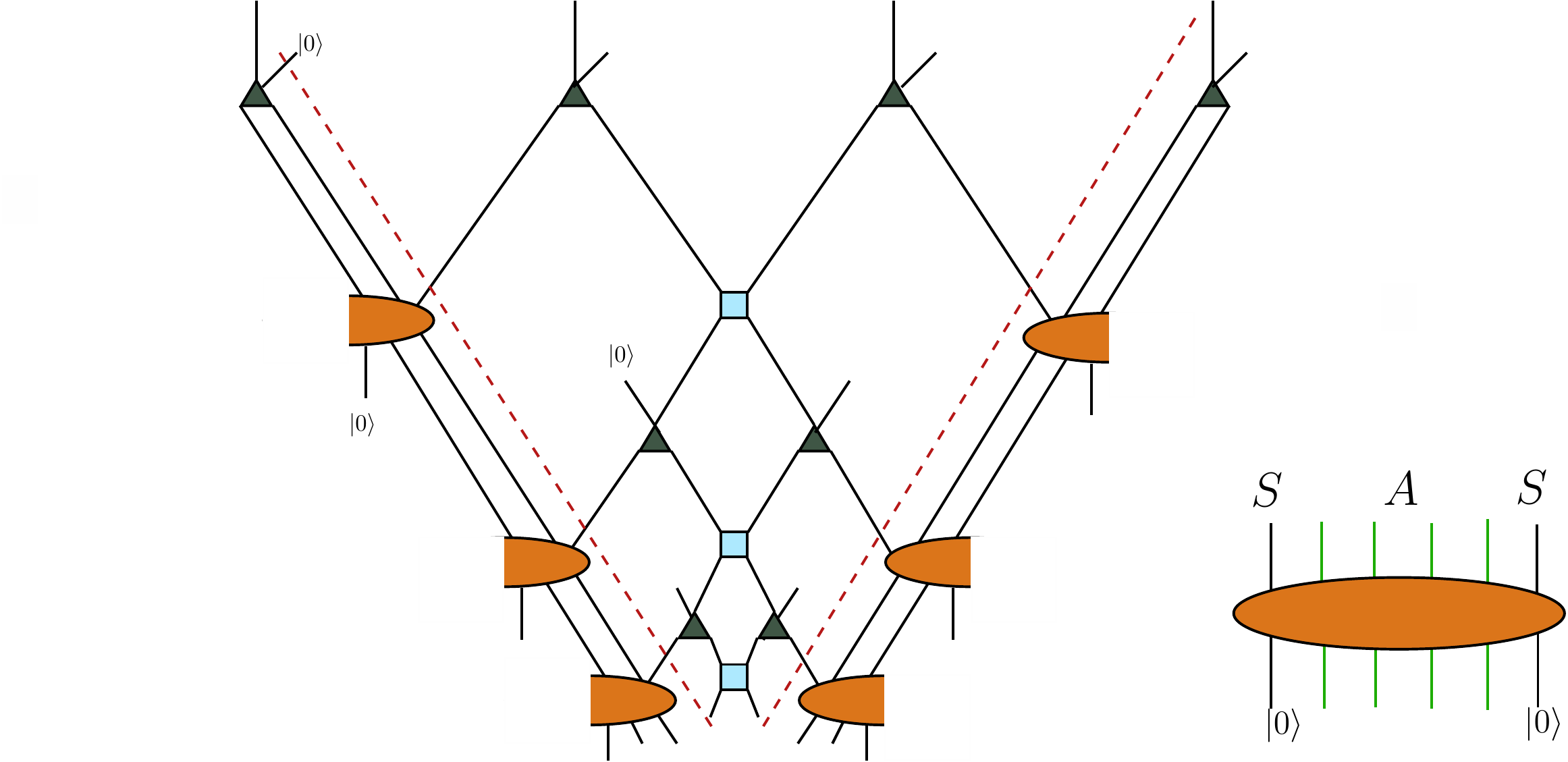}
\caption{The strong complementarian version of MERA that describes a static patch for a local observer with horizon degrees of freedom. The future direction points downward in the fine-graining direction. Dashed red lines demarcate the interior of the de~Sitter static patch. The combined system, including a constant number of ancillae, evolves unitarily. The horizon degrees of freedom at each time step are acted upon by a single recovery tensor (orange ellipse), which serves as a map that distills the same ancillary state (represented by $\ket{0}$ in the figure) that is entangled in the interior at the horizon. (Half-ellipses on opposite sides of the tensor network are identified.) The ancillary system is denoted by $S$ while the horizon degrees of freedom are denoted by $A$. }
\label{fig:SCMERA}
\end{figure}

To turn the network into a description of a single-patch universe, we propose modifying the MERA circuit as follows.
First, choose any single static patch in the MERA (cf. \Fig{fig:dS-MERA}).
At $t=0$, we identify the degrees of freedom inside a static patch as interior degrees of freedom living in the Hilbert space $\mathcal{H}_{\mrm{static}}$.
The remaining exterior degrees of freedom in the other patch can now be identified with the horizon within the Hilbert space $\mathcal{H}_{\mrm{horizon}}$, with $\dim \Hil_\mrm{static} = \dim \Hil_\mrm{horizon} < \infty$.
For a local picture, we preserve the circuit structure for the static patch interior, but now we introduce separate circuit dynamics for $\Hil_\mrm{horizon}$, as shown in \Fig{fig:SCMERA}. In particular, a recovery tensor (indicated by the ellipse) acts to extract ancilla states at the horizon. 
Because the interior network is unchanged, the previous cosmic no-hair result about the interior state continues to hold. 

This circuit structure constrains the action of the recovery tensor that acts on $\Hil_\mrm{horizon}$ in \Fig{fig:SCMERA} if we demand unitary evolution.
At each time step, new ancillae are mixed with the interior via the action of the isometries (triangular tensors), and then some information will flow to the horizon and become inaccessible to any interior observer via the action of the disentanglers (square tensors).
To be consistent with the literature, label the Hilbert space of the ancillae by $S$, the static patch Hilbert space by $E$ (i.e., $\Hil_\mrm{static} \equiv E$), and the horizon Hilbert space by $A$ (i.e. $\Hil_\mrm{horizon} \equiv A$).
If it is always the same ancillary state $\sigma_S$ (which we have simply taken to be $\sigma_S = \ketbra{0}{0}_S$ throughout) that gets mixed in via the isometries, then in order to have consistent unitary evolution, it must be that the recovery tensor, which acts on $AS$, must spit out a state of the form $\rho^{\prime\prime}_A \otimes \sigma_S$.
Put another way, if at every time step we re-introduce a fresh ``copy'' of the ancillary state $\sigma_S$, then unitarity in each time step demands that $\sigma_S$ be restored after evolving forward in time.
(Alternatively we could drop the requirement of unitary evolution; we will return to this possibility at the end of this section.)
We call such a circuit for the local picture the Strong Complementarity MERA (SCMERA).
The usual global picture can be easily restored by allowing ourselves more ancillary degrees of freedom and replacing the horizon tensors with the usual MERA circuit. 
As a result, the local and global pictures are related by some global unitary transformation that act on the extended set of ancillae.

\begin{figure}
 \centering
 \includegraphics[width=0.45\textwidth]{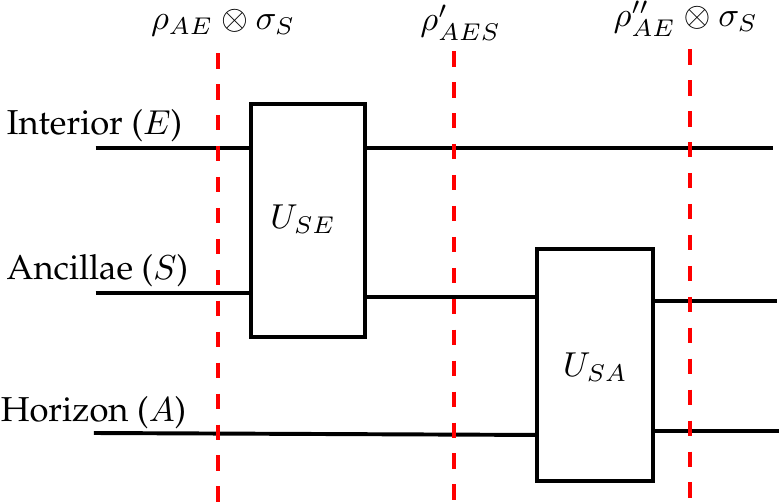}
 \caption{Each time step of SCMERA can be condensed into a circuit diagram. The dashed lines mark the resulting quantum state at the end of a subprocess. In the case where the MERA global state is pure, which is the case we consider here, it follows that $\rho_{SEA}$, $\rho'_{SEA}$, and $\rho''_{SEA}$ are all pure states.  }
 \label{fig:rec_circuit}
\end{figure}

Let us ask whether it is possible to have a circuit with the tensor structure in \Fig{fig:SCMERA} that spits out the state $\sigma_S$ at every time step.
To answer this question, it is useful to analyze the SCMERA circuit from the perspective of recovery maps.
At each time step of SCMERA, we can describe the quantum process by
\begin{equation}
 \rho_{AES}=\rho_{AE}\otimes\sigma_S \xrightarrow[]{U_{SE}\otimes I_A} \rho'_{AES}\xrightarrow[]{U_{SA}\otimes I_E} \rho''_{AES}= \rho''_{AE}\otimes\sigma_{S},
\end{equation}
as shown in the quantum circuit diagram in \Fig{fig:rec_circuit}. 
$U_{SE}$ corresponds to the isometries that entangle the ancillae and the interior degrees of freedom, as well as the disentanglers, while $U_{SA}$ acts on the horizon.
Since $U_{SA}$, which corresponds to the elliptical orange tensor in Figure \ref{fig:SCMERA}, must recover the state $\sigma_S$, we call it the recovery tensor.

Although the existence of such a recovery tensor is not always guaranteed, we can examine the necessary conditions that these tensors and states must satisfy to allow such a recovery operation.
For instance, if the ancillary qudit is always initialized in a fixed vector, e.g., $\ket{0}_S$, or more generally is always chosen from some fixed subspace of $S$, then one can derive necessary conditions for recoverability by appealing to results from quantum error correction. 

To understand the recoverability of the ancillary state, we first consider the action of $U_{SE}$ as a quantum channel on $S$, $\mathcal{N}_{\rho_{SE}}:S(\mathcal{H}_S)\rightarrow S(\mathcal{H}_S)$.
This is always possible because the initial state is uncorrelated across $S$ and $E$.
However, because the state in $E$ is in principle arbitrary (and certainly will change at each time step if the mapping is not at a fixed point), the channel can depend on the input $\rho_{SE}$.
Likewise, the recovery tensor will not remain fixed at every time step.
This is what we mean when we say that the SCMERA describes evolution that is generated by a time-dependent Hamiltonian; the recovery tensor will change at every time step if it must recover $\sigma_S$ exactly.

Given such a channel and knowledge of the fixed ancillary state $\sigma$, there always exists a process in the reduced system $S$ that recovers $\sigma$. 
Let $\sigma\geq 0$ be the known state in which the ancillary system is initialized.
In general, there exists a completely positive trace preserving (CPTP) recovery map $R$ such that $R\circ \mathcal{N}_{\rho_{SE}}(\sigma')=\sigma'$ for all $\sigma'$ if and only if the monotonicity condition is saturated \cite{Petz1986,Petz1988,Petz2003,Winter:2015}: 
\begin{equation}
 D(\sigma' \Vert \sigma) = D(\mathcal{N}_{\rho_{SE}}(\sigma') \Vert \mathcal{N}_{\rho_{SE}}(\sigma)),
 \label{eqn:monotonicity}
\end{equation}
where $D(\sigma' \Vert \sigma) $ is the relative entropy between $\sigma'$ and $\sigma$.
In particular, $\sigma$ is always recoverable because the monotonicity condition is trivially saturated when $\sigma'=\sigma$. 
For the finite-dimensional case, one can construct an explicit Petz recovery map $P$ that will always recover $\sigma'$: 
\begin{equation}
 P_{\sigma,\mathcal{N}_{\rho_{SE}}}: X \mapsto \sigma^{1/2}\mathcal{N}_{\rho_{SE}}^{\dagger}(\mathcal{N}_{\rho_{SE}}(\sigma)^{-1/2}X\mathcal{N}_{\rho_{SE}}(\sigma)^{-1/2})\sigma^{1/2}.
\end{equation}
Since we here consider the trivial case where $\sigma'=\sigma$, the Petz map can always recover $\sigma$. 

Unfortunately, in the case of interest here the existence of a Petz recovery map does not lead us to the sought-after unitary $U_{SA}$, since the Petz map doesn't necessarily take the form of a partial trace $\Tr_A( U_{SA} \, \rho^\prime_{SA} \, U_{SA}^\dagger )$.
Indeed, we can in fact argue that the Petz map \emph{cannot} identically be the map $\Tr_A( U_{SA} \, \rho^\prime_{SA} \, U_{SA}^\dagger )$.
This latter recovery map cannot be CP over the set of all density operators if $A, S,E$ are in an entangled state, which will generally be the case\footnote{Even if one fixes a particular input at $t=0$ to be a product state, entanglement will still be generated at a later time. This is because $S,E$ generically become entangled after the isometry.}, whereas the Petz map is CP by construction. So while $U_{SA}$ may exist, it cannot be found in this way.

In light of this difficulty, a different line of attack is to use the given unitary structure of the SCMERA as a starting point and see whether recovery can be engineered.
This amounts to interpreting recovery as an instance of quantum error correction that protects against deletion of $E$.
Think of the state $\sigma_S$ that the ancillae are initialized in as an encoded message.
At any given time step, the message is encoded into the combined $SEA$ system by entangling it with $EA$.
A part of the system, $E$, subsequently becomes inaccessible to us.
We then wish to recover the encoded message by acting on the reduced $SA$ system only with $U_{SA}$.
If this is to be possible, then the allowed interactions $U_{SE}$ are constrained.
(This picture is reminiscent of quantum secret sharing.)

Since $\sigma_S$ is the message that we want to recover and since we discard $E$, here $\mathcal{N}_{\rho_{SE}}$ is essentially a noisy channel, which we suppose takes on a particular Kraus form,
\begin{equation}
\mathcal{N}_{\rho_{SE}}: X \mapsto \sum_\mu N_{\mu} X N_{\mu}^{\dagger},
\end{equation}
for a given initial state $\rho_{SE}$\footnote{Recall that any trace-preserving channel on a reduced system can be written using a (potentially input-dependent) set of Kraus operators $\{N_\mu\}$, where $\sum_\mu N^{\dagger}_\mu N_\mu=I$ \cite{Rivas:2012}.}.
In this context, in order for a recovery map $R$ to exist, the Kraus operators $N_\mu$ must obey the following necessary and sufficient condition \cite{Nielsen:2011:QCQ:1972505}.
For the sake of generality, suppose that instead of wanting to recover a fixed state $\ket{0}_S$, the encoded message was chosen from a fixed subspace of $S$ that has an orthonormal basis $\{ \ket{\phi_i}_S \}$ (the specific case for SCMERA corresponds to there only being one basis vector, namely, $\ket{0}_S$).
Then, the Kraus operators must obey the Knill-Laflamme condition,
\begin{equation}
 \bra{\phi_i} N_{\mu}^{\dagger}N_{\nu}\ket{\phi_j} = C_{\mu\nu} \delta_{ij} \, ,
 \label{eqn:QECC}
\end{equation}
where $C_{\mu\nu}$ is a Hermitian matrix.
This condition places a constraint on what $U_{SE}$ are allowed.

In the case of a single fixed state $\ket{0}_S$, the condition above is trivially satisfied, and so recovery is always possible.
However, here as well it is not guaranteed whether there is a quantum error correcting code (QECC) on the whole $SEA$ system that is consistent with SCMERA such that the ancillary state can always be recovered on the $SA$ subsystem on the horizon.
We do not know whether such a code exists, but it would have to satisfy certain requirements that we now explore.

In the case where the ancillary qudit is fixed to be a particular state, the code subspace is 1-dimensional.
An implementation that allows one to decode the message may be possible to realize with the help of a $k=0$ code\footnote{The properties of a quantum error correcting code on qudits of dimension $\chi$ are often abbreviated by the notation $[\![n,k,d]\!]$ where $n$ is the block size, $k$ is the number of encoded qudits, and $d$ the code distance. For $k=0$, the  $\chi^k$-dimensional code subspace is precisely one dimensional.}.
(See \cite{Nielsen:2011:QCQ:1972505} for a detailed review.)
For a binary MERA, in which the interior, horizon, and ancillary Hilbert spaces are altogether comprised of $8$ qudits, a satisfactory encoding would require a $[\![8,k,d]\!]$ code, where $k=0$ if the ancillary states are always fixed to be $\ket{0}_S$.
Because 2 qudits are effectively erased in discarding the interior (i.e., a known erasure location), the distance of the code must satisfy $d\geq t+1$ with $t= 2$.
As a zeroth order check, we see that this requirement is consistent with the quantum Singleton bound
\begin{equation}
 n-k\geq 2(d-1)
\end{equation}
for $3\leq d \leq 5$ with $k=0$.
Also note that, while we mainly consider the case where $k=0$, larger code spaces with $k > 0$ (i.e., a situation where the ancillary state is chosen among several options at each step) are not ruled out.
For example, a hypothetical tensor network that encodes $k = 2$ qudits worth of information could realize a QECC with $d = 3, 4$.
We note that there exist binary codes that are compatible with our requirements on $n$, $k$, and $d$, for example, the $[\![8,3,3]\!]$ code (see section 7.12.3 in \cite{preskillLecture}), and presumably there also exist codes for qudit systems; however, we are unaware of their specific forms, and much less whether or not they are compatible with the tensor structure of SCMERA.

In summary, by interpreting SCMERA as a recovery operation or an error correcting code, we identify several necessary but generally insufficient criteria that the SCMERA circuit must meet.
Note, however, that failure to meet these criteria cannot rule out strong complementarity, but it can rule out SCMERA as a model.

Finally, we elaborate a bit more on the unitarity of the proposed SCMERA circuit. The overall SCMERA tensor network can be understood as a circuit by including the ancillary degrees of freedom, $S$. In the case of perfect recovery of the ancillary state on the horizon, the ancillary state that was added in the interior can be discarded from the horizon at the end of the computation in each time step so that the total size of Hilbert space remains constant throughout. Alternatively, we can also understand the adding-and-discarding process as recycling the ancillary degrees of freedom at each step. It is clear in this sense that we have a unitary process on the same finite-dimensional Hilbert space. However, note that the unitary recovery mapping on the horizon need not recover the ancillary state perfectly. In fact, a universal (i.e., constant in time) unitary recovery map applied to every time step cannot in general achieve perfect recovery.
 In this case, recycling of the approximately recovered ancillary qudit will lead to information backflow into the static patch interior, which in turn leads to Poincar\'e recurrences.  Discarding such ancillary qudits on the horizon avoids recurrences even when using a universal recovery map, but breaks unitarity. If we demand perfect recovery of the ancillary qudit, then the unitary evolution is necessarily time-dependent. It is, however, unclear if such time-dependence is only limited to swapping operations on the horizon.

\section{Circuit Complexity and de~Sitter Action}

In AdS/CFT, the ``complexity equals action'' proposal \cite{Brown:2015bva} suggests that the complexity of a CFT thermofield double state as it evolves in time is proportional to the Einstein-Hilbert (EH) action of a region of the bulk known as the Wheeler-De~Witt patch.
Explicitly, $\mathcal{C} = q S_{\rm EH}$, where the proportionality constant is calculated to be $q=1/\pi \hbar$.
Similarly, here we can show that complexity, calculated using the MERA circuit, scales in the same way as the corresponding spacetime action in de Sitter space.

For a given MERA-like circuit that is translationally and scale invariant, it is possible to estimate its complexity by choosing a reference state and gate set.
It is natural to choose the reference state to be the initial state of dS/MERA, which we write as $|\Psi(t=0)\rangle = |\psi\rangle\otimes |\phi\rangle^{\otimes N}$.
$|\Psi\rangle$ consists of the initial entangled component $|\psi\rangle$ which encodes the entanglement information needed to reconstruct the de~Sitter spatial geometry at $t=0$, and $|\phi\rangle^{\otimes N}$ denotes all the ancillary degrees of freedom that will later get entangled up to some time $t=T$.
Here, because we only consider bounds on complexity, the estimate won't depend on the particular form of $|\psi\rangle$; we can take it to be an arbitrary state that lives on the initial few sites of the MERA at $t=0$. 

We obtain a straightforward estimate of complexity if we choose a reference gate set that corresponds to the exact disentanglers and isometries, $\{ U, V\}$, that were used to build the MERA circuit.
For a $k$-nary MERA, suppose that $U,V$ are $k$-local and denote the total number of ancillae that get entangled up to time $t\leq T$ by
\begin{equation}
N(T) = \sum_{j=0}^{T} k^j.
\end{equation}
It then follows that for any non-trivially entangled state $|\Psi(T)\rangle$, where none of the qudits in $|\Psi(T)\rangle$ can be written as a product state between the qudit and its complement\footnote{For example, this is expected for a CFT vacuum state.}, a lower bound on its complexity $\mathcal{C}(T)$ is proportional to $N(T)$. %This is true regardless of the exact circuit construction. 
This is because, even using an optimal circuit that could potentially be more efficient than the MERA, it takes at least $N(T)/k$ $k$-local gates to even minimally entangle all of the product ancillae.
The actual complexity to create the state with the correct entanglement structure at $t=T$ is therefore strictly lower-bounded.
In addition, the MERA circuit itself that constructs the state $|\Psi(T)\rangle$ constitutes a trivial complexity upper bound.
Hence, for generic scale and translationally invariant MERA in arbitrary dimensions with $k$-local disentanglers and isometries, the complexity satisfies
\begin{equation}
 C_0 N(T)  \leq \mathcal{C}(T)\leq C_1 N(T),
% C_0\sum_{j=0}^{T} k^j \leq \mathcal{C}(T)\leq C_1 \sum_{j=0}^{T} k^j,
\end{equation}
where $C_1>C_0$ are  order-unity numbers that depend on the specific circuit construction. For the (1+1)-dimensional binary MERA shown, $C_0=4$ and $C_1=8$.
Choosing a different reference gate set would give different coefficients $C_0$ and $C_1$, but the exponential dependence on $T$ would remain unchanged.

An important distinction from the usual holographic complexity proposal \cite{Brown:2015bva} is the lack of a boundary theory, and hence a notion of bulk-boundary duality. Similarly, the proposal also differs from \cite{Reynolds:2017lwq}, where the complexity of the state on the de~Sitter boundary is compared to the action or volume of a holographic asymptotically anti-de~Sitter bulk.
Because only the de~Sitter bulk is present, we test a bulk complexity-action (volume) proposal by directly comparing the circuit complexity of MERA, which is conjectured to describe de Sitter spacetime, to the Einstein-Hilbert action (spacetime volume) of the same region in de Sitter. 

The Einstein-Hilbert action of the portion of de Sitter spacetime covered by the global time interval $0\leq t\leq T$ in $D$ dimensions is given by
\begin{align*}
 S_{\rm EH} &= \frac{1}{16\pi G} \int_0^T dt \, \int d\Omega_{D-1} \, \sqrt{-g} R \\[2mm]
&= \frac{R \ell_{\rm dS}^{D} \mathcal{S}_{D-1}}{16 \pi G} \int_0^T dt \, \cosh^{D-1} t \\[2mm]
 &= \frac{R \ell_{\rm dS}^{D} \mathcal{S}_{D-1}}{16 \pi G} \frac{1}{(D-1)2^{D-1}} e^{(D-1)T} ~ + ~\rm subleading,
\end{align*}
where $R=D(D-1)/\ell_{\rm dS}^2 = 2D\Lambda /(D-2)$ is the Ricci curvature for de~Sitter space with cosmological constant $\Lambda$  and $\mathcal{S}_{D-1}$ is the volume of the $(D-1)$-sphere.
We see that the scaling behavior is indeed consistent with the circuit complexity computed above, and the action satisfies the complexity bound for some appropriate choice of constant $q$.
Note that each tensor in the MERA is mapped to a proper volume in de Sitter \cite{Czech:2015kbp}. Therefore, comparison of other spacetime regions would yield a similar conclusion. It cannot differentiate the complexity = volume versus complexity = action proposal, because the  constant Ricci curvature in de~Sitter space only changes $q$ by a constant factor. 

The proportionality constant between complexity and action depends on the choice of gate set, and differs from $q=1/\pi \hbar$ in the original proposal. See \cite{Jefferson:2017sdb,Chapman:2017rqy} for similar conclusions from more detailed studies in the context of quantum field theory. Interestingly, assuming the validity of the conjecture, the $(\ell_{\rm dS}/\ell_{\rm pl})^{D-2}$ scaling behavior in the action may suggest that the complexity of a correct circuit with sub-Hubble features should approximately scale as the horizon area (recall that $R$ scales like $\ell_\mrm{dS}^{-2}$). In the case of the MERA, this is encoded in the otherwise arbitrary choice of $q$, because the network structure is not sensitive to $\ell_{\rm dS}/\ell_{\rm pl}$.

\section{Discussion}

Discretizing de~Sitter spacetime using the MERA seems to provide some interesting interpretations, in particular in terms of giving a natural information-theoretic reason for cosmic no-hair, constraining de~Sitter complementarity, and giving the de~Sitter action an information-theoretic interpretation.
It would be interesting to ask what other consequences thinking of de~Sitter spacetime in a tensor network/information-theoretic way could provide.
For example, would a different tensor network discretization be more natural for answering other questions, or is the choice of tensor network discretization fixed by the spacetime metric one is attempting to duplicate?
If so, are there other natural spacetimes (Lorentzian or Euclidean), for which different tensor networks might provide insights into open problems?

The MERA is naturally suited to describing de~Sitter spacetime on super-Hubble scales, since structure within a horizon volume is not resolved.
The state within a patch can nevertheless be encoded in the tensors inside the horizon, and perturbations of such a state in the de~Sitter background can be initialized in the MERA input state.
The cosmic no-hair result is then the fact that such perturbations flow to a fixed-point of the evolution superoperator within a patch.

Another limitation of this de~Sitter-MERA correspondence is that it clearly breaks the rotational symmetry of spacelike sections of de~Sitter;
a binary MERA that corresponds to (1+1)-dimensional de~Sitter spacetime picks out four preferred causal patches, or equivalently, fixes the cardinal directions on the circle.
It also breaks boost symmetry in that the MERA fixes a preferred global $t=0$ slice.
To this end, hyperinvariant tensor networks may be an interesting improvement on the MERA \cite{Evenbly:2017}.
Hyperinvariant tensor networks were introduced to address, among other issues, a similar problem for AdS-MERA correspondences that the MERA picks out a preferred center point of the hyperbolic plane.
In a hyperinvariant tensor network, any node in the tensor network can be taken to be the ``center'' of the hyperbolic plane, thus restoring a significant amount of symmetry.
Since the radial direction in AdS corresponds to the renormalization direction of the MERA, which here corresponds to the timelike direction of de~Sitter, a hyperinvariant tensor-network/de~Sitter correspondence would likely no longer fix a preferred global $t=0$ slice.
Instead, the effective causal cone of any pair of adjacent nodes could be used to define a de~Sitter static patch.

It would be interesting to push the present analysis beyond a strict de~Sitter background.
For example, it should be possible to adapt the tensor network to allow for bubble nucleation and eternal inflation.
A classical variant of this was already considered in \cite{Harlow:2011az}, and it would be useful to further investigate the evolution of quantum states using the kind of methods explored here.

\section*{Acknowledgments}

We would like to thank Aleksander Kubica, Yasunori Nomura, Jason Pollack, and Grant Remmen for helpful discussions.
This material is based upon work supported by the U.S. Department of Energy, Office of Science, Office of High Energy Physics, under Award Number DE-SC0011632, as well as by the Walter Burke Institute for Theoretical Physics at Caltech and the Foundational Questions Institute.

\appendix
\section{Stationary causal cones of the MERA}
\label{app:stationary}

Given a $k$-nary MERA, in which the number of sites in each layer increases $k$-fold with every fine-graining step, what is the number of sites per layer of a stationary causal cone?

\begin{figure}[ht]
\centering
\subfloat[]{
\includegraphics[scale=1,width=0.3\textwidth]{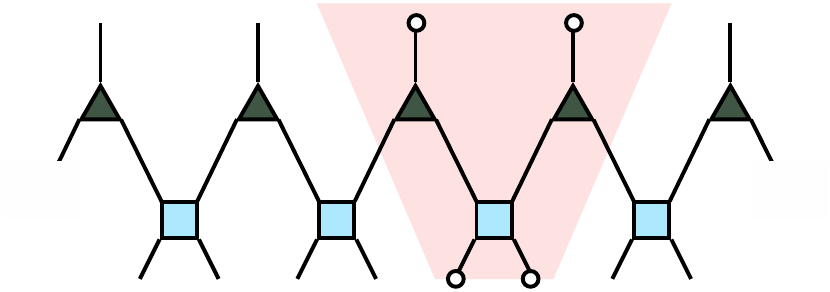}
}
\subfloat[]{
\includegraphics[scale=1,width=0.3\textwidth]{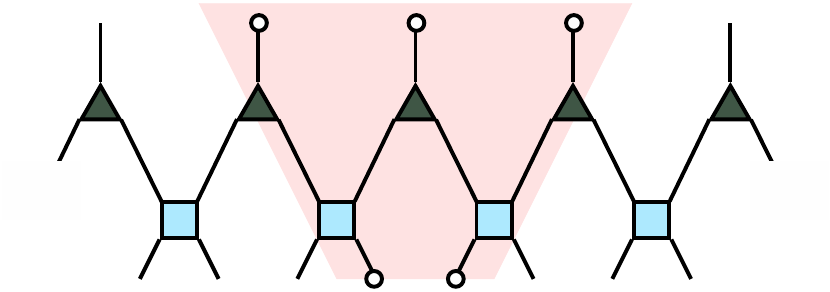}
}
\subfloat[]{
\includegraphics[scale=1,width=0.3\textwidth]{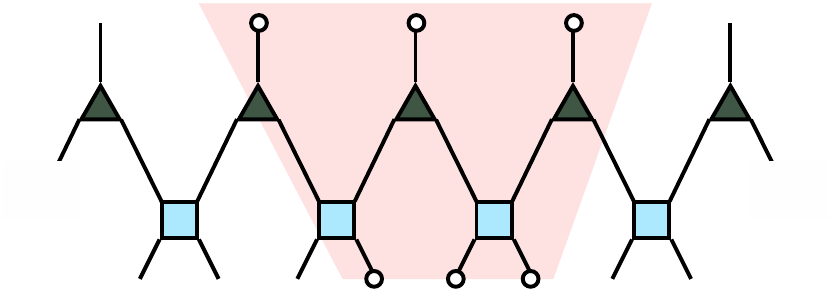}
}
\caption{Ways in which a minimal-width causal cone can propagate between layers in a binary MERA. (a) $2 \rightarrow 2$, (b) $2 \rightarrow 3$, (c) $3 \rightarrow 3$. }
\label{fig:binaryx-x}
\end{figure}

\begin{eg}
Consider a binary MERA as in \Fig{fig:binaryMERA}.
Within the MERA, consider a set of sites at some layer and draw their causal cone in the coarse-graining direction.
If the smallest simply-connected region that contains all of the initial sites is made up of $L$ sites, then after $\sim \log_2 L$ steps in the coarse-graining direction, the causal cone will contain 2 or 3 sites \cite{Evenbly2014}.
Once the cone at some layer contains 2 or 3 sites, \Fig{fig:binaryx-x} illustrates how the width of the causal cone can evolve under further coarse-graining.
Notice that if the cone contains 2 sites at some layer, then it is possible for the next layer to have either 2 or 3 sites, but if a given layer contains 3 sites, then all subsequent layers will contain 3 sites.
Therefore, a stationary causal cone having the same width at every layer can only have 2 sites per layer or 3 sites per layer.
In particular, only the stationary causal cone with 2 sites per layer is left/right-symmetric in a binary MERA.
\end{eg}

\begin{figure}[ht]
\centering
\subfloat[]{
\includegraphics[scale=0.7,width=0.18\textwidth]{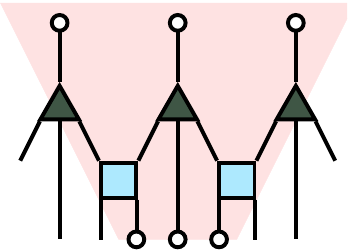}
} ~~~~~
\subfloat[]{
\includegraphics[scale=0.7,width=0.18\textwidth]{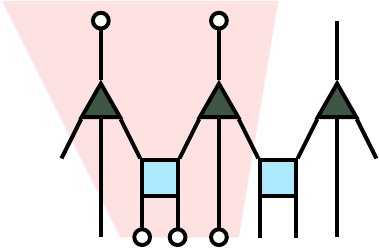}
} \\
\subfloat[]{
\includegraphics[scale=0.7,width=0.18\textwidth]{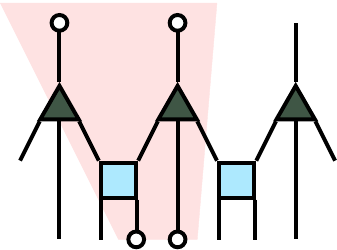}
} ~~~~~
\subfloat[]{
\includegraphics[scale=0.7,width=0.18\textwidth]{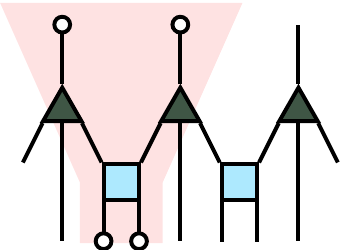}
}
\caption{Ways in which a minimal-width causal cone can propagate between layers in a ternary MERA. (a) $3 \rightarrow 3$, (b) $3 \rightarrow 2$, (c) $2 \rightarrow 2$, first instance, (d) $2 \rightarrow 2$, second instance. }
\label{fig:ternaryx-x}
\end{figure}

\begin{eg}
Consider a ternary MERA as in \Fig{fig:ternaryMERA}.
Similarly, the causal cone of any given collection of sites will contain 2 or 3 sites after $\sim \log_3 L$ steps in the coarse-graining direction.
If the cone contains 3 sites at some layer, then it is possible for the next layer to have either 2 or 3 sites, but if a given layer contains 2 sites, then all subsequent layers will contain 2 sites (\Fig{fig:ternaryx-x}).
Therefore, a stationary causal cone having the same width at every layer can only have 2 sites per layer or 3 sites per layer.
Here, only the stationary causal cone with 3 sites per layer is left/right-symmetric in a ternary MERA.
\end{eg}

The case of a general $k$-nary MERA follows straightforwardly from the two examples above:

\begin{prop}
A stationary causal cone having the same width at every layer in a homogeneous $k$-nary MERA has 2 or 3 sites per layer.
\end{prop}

\begin{figure}
\centering
\includegraphics[scale=1]{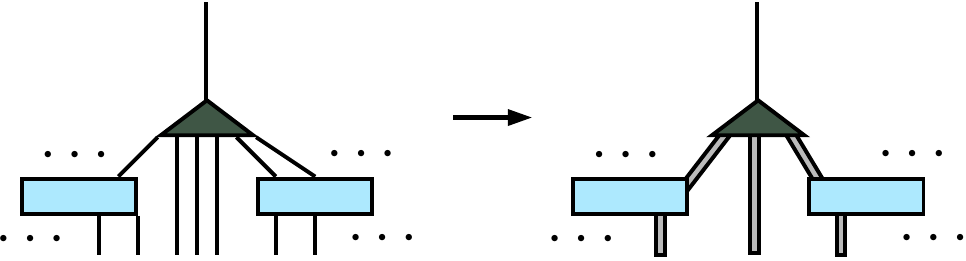}
\caption{Legs in an arbitrary MERA can be blocked together. In this way, that the causal structure matches that of a binary or ternary MERA becomes apparent.}
\label{fig:blocking}
\end{figure}

\pf
Given some homogeneous $k$-nary MERA with any arrangement of disentanglers and isometries, all of the legs in the tensor network can be blocked together to form composite legs so that the network takes the form of a binary or ternary MERA, as illustrated in \Fig{fig:blocking}, whence the proposition follows from the examples above.
\eop

\section{Higher-dimensional generalizations}
\label{app:MOREDIMS}

\begin{figure}[h!]
\centering
\includegraphics[width=\textwidth]{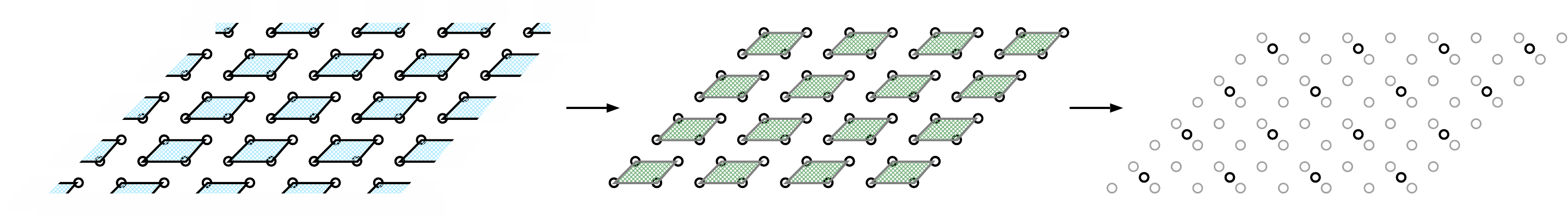}
\caption{A 2D MERA. In a single coarse-graining step, blocks of 4 sites are acted on by a disentangler (blue), then blocks of 4 sites that are displaced from the last set of blocks are acted on by an isometry (green), reducing the number of sites by a factor of $4$.}
\label{fig:2DMERA}
\end{figure}

Consider a $d$-dimensional MERA, where each layer is a hypercubic $d$-dimensional lattice.
Here, the MERA is $k$-nary when each site in one layer gives rise to $k^d$ sites in the next layer (see \Fig{fig:2DMERA}). 
The global MERA-de~Sitter correspondence does not carry through in this case, simply because, on the de Sitter side, there is no way to latticize the $d$-sphere using a regular hypercubic lattice that is self-similar under fine-graining (although see \cite{Evenbly:2008} for a generalization to 2 dimensions).

This is not to say that a generalization to higher dimension is impossible.
One could consider a different tiling of global de Sitter that preserves uniformity and is self-similar under some refinement operation.
For example, on a 2-sphere, regular or semi-regular tilings are possible using triangularizations, but these different tilings would necessarily require some sort of variation on the MERA tensor network.
To the best of our knowledge, such generalizations are still unexplored.

On the other hand, one could still study the correspondence between de~Sitter and a hypercubic MERA by restricting one's attention to only a single static patch.
In this scenario, it is consistent to think of the MERA as defining a superoperator which maps the state on $m^d$ sites of a given slice of a single static patch to the next slice.
(Remember, the number of sites per horizon volume, i.e., per slice of the static patch, remains constant.)
Therefore, the usual unmodified MERA may still be useful for understanding local aspects of de Sitter quantum gravity in higher dimensions.

\bibliographystyle{JHEP}
\bibliography{dS-MERA}

\end{document}